\documentclass{article}
\usepackage{graphicx}
\newcommand{\blitza}{\text{\usefont{U}{ulsy}{m}{n}\symbol{'011}}}
\usepackage{fullpage}
\usepackage{graphicx,framed}
\usepackage{amsmath,amssymb,latexsym,amsopn,amsfonts,amsthm}
\usepackage[boxed,ruled,vlined,linesnumbered,noresetcount]{algorithm2e}
\makeatletter
\newcommand{\nosemic}{\renewcommand{\@endalgocfline}{\relax}}
\newcommand{\dosemic}{\renewcommand{\@endalgocfline}{\algocf@endline}}
\newcommand{\pushline}{\Indp}
\newcommand{\popline}{\Indm\dosemic}
\let\oldnl\nl
\newcommand{\nonl}{\renewcommand{\nl}{\let\nl\oldnl}}
\makeatother
\newcommand{\bZ}{{Z}\xspace} 
\newcommand\bull{{\operatorname{-\xspace}}}

\usepackage{amsmath,amssymb,amsfonts}
\usepackage{graphicx}
\usepackage{textcomp}
\usepackage{xcolor}
\def\BibTeX{{\rm B\kern-.05em{\sc i\kern-.025em b}\kern-.08em
	T\kern-.1667em\lower.7ex\hbox{E}\kern-.125emX}}

\newcommand{\EMS}[1]{[\textcolor{red}{EMS: #1}]} 
\newcommand{\emsT}[1]{\textcolor{black}{#1}}
\definecolor{oskar_green}{rgb}{0.0, 0.5, 0.0}

\newcommand{\algSize}{normalsize} 
\newcommand{\algSizeSmall}{normalsize} 

\newcommand{\FF}{\vspace*{\medskipamount}}

\newcommand{\bigO}{\mathcal{O}\xspace}
\newcommand{\remove}[1]{}
\newcommand{\tr}[1]{\textcolor{black}{#1}}
\newcommand{\ea}[1]{\textcolor{black}{#1}}

\newcommand{\nextQueryJ}{\mathit{nextQuery}\mathrm{J}\xspace}
\newcommand{\nextQuery}{\mathit{nextQuery}\xspace}

\newcommand{\maxSeq}{\mathit{maxSeq}\xspace}

\newcommand{\nonActive}{\mathit{allHaveTerminated}\xspace}
\newcommand{\done}{\mathit{result}\xspace}

\newcommand{\Figure}{Fig.\xspace}
\newcommand{\minReady}{\mathit{minReady}\xspace}
\newcommand{\maxReady}{\mathit{maxReady}\xspace}
\newcommand{\bulkRead}{\mathit{bulkRead}\xspace}
\newcommand{\getSeq}{\mathit{getSeq}\xspace}
\newcommand{\7}{{3}} 
\newcommand{\6}{{2}} 
\newcommand{\5}{{1}} 
\newcommand{\SYNC}{\mathrm{SYNC}\xspace}
\newcommand{\SYNCack}{\mathrm{SYNCack}\xspace}
\newcommand{\CS}{\mathit{CS}\xspace}
\newcommand{\Sset}{\mathit{actCS}\xspace}

\newcommand{\exceed}{\mathit{needFlush}}

\newcommand{\etal}{\emph{et al.}\xspace}
\newcommand{\eg}{\emph{e.g.,}\xspace}

\newcommand{\ie}{\emph{i.e.,}\xspace}
\newcommand{\Ie}{\emph{I.e.,}\xspace}

\newcommand{\xS}{\mathit{obsS}\xspace}

\remove{
\newcommand{\nextQueryJ}{\mathsf{nextQuery}\mathrm{J}\xspace}
\newcommand{\nextQuery}{\mathsf{nextQuery}\xspace}

\newcommand{\maxSeq}{\mathsf{maxSeq}\xspace}

\newcommand{\nonActive}{\mathsf{allHaveTerminated}\xspace}
\newcommand{\done}{\mathsf{result}\xspace}

\newcommand{\Figure}{Fig.\xspace}
\newcommand{\minReady}{\mathsf{minReady}\xspace}
\newcommand{\maxReady}{\mathsf{maxReady}\xspace}
\newcommand{\bulkRead}{\mathsf{bulkRead}\xspace}
\newcommand{\getSeq}{\mathsf{getSeq}\xspace}
\newcommand{\7}{{3}} 
\newcommand{\6}{{2}} 
\newcommand{\5}{{1}} 
\newcommand{\SYNC}{\mathrm{SYNC}\xspace}
\newcommand{\SYNCack}{\mathrm{SYNCack}\xspace}
\newcommand{\CS}{\mathit{CS}\xspace}
\newcommand{\Sset}{\mathsf{actCS}\xspace}

\newcommand{\exceed}{\mathit{needFlush}}

\newcommand{\etal}{\emph{et al.}\xspace}
\newcommand{\eg}{\emph{e.g.,}\xspace}

\newcommand{\ie}{\emph{i.e.,}\xspace}
\newcommand{\Ie}{\emph{I.e.,}\xspace}

\newcommand{\xS}{\mathit{obsS}\xspace}

} 

\newtheorem{theorem}{Theorem}[section]
\newtheorem{lemma}[section]{Lemma}
\newtheorem{definition}{Definition}[section]

\newtheorem{assumption}{Assumption}[section]
\newtheorem{claimA}[section]{Claim}
\newcommand{\true}{\mathsf{True}\xspace}

\newcommand{\false}{\mathsf{False}\xspace}

\newenvironment{claimProof}{\par\noindent\textbf{Proof of Claim  \clmcnt\space}}{\hfill $\Box_{Claim ~ \clmcnt}$}

\newenvironment{lemmaProof}{\par\noindent\textbf{Proof of Lemma  \lemcnt\space}}{\hfill $\Box_{Lemma ~ \lemcnt}$}

\newenvironment{theoremProof}{\par\noindent\textbf{Proof of Theorem  \thmcnt\space}}{\hfill $\Box_{Theorem ~ \thmcnt}$}

\newcommand{\clmcnt}{0}
\newcommand{\lemcnt}{0}
\newcommand{\thmcnt}{0}

\newcommand{\sP}{\mathcal{P}\xspace}

\newcommand{\capacity}{\mathsf{channelCapacity}\xspace}

\newcommand{\Correct}{\mathit{Correct}\xspace}

\newcommand{\Section}[1]{\section{#1}}
\newcommand{\Subsection}[1]{\subsection{#1}}
\newcommand{\Subsubsection}[1]{\subsubsection{#1.}}
\newcommand{\Subsubsubsection}[1]{\paragraph{#1.}}

\usepackage{enumitem}

\begin{document}
\title{Self-stabilizing Total-order Broadcast}

%
%

\author{Oskar Lundstr{\"{o}}m~\footnote{Chalmers Univ. Tech., Sweden \texttt{\{osklunds@student.,elad@\}chalmers.se}} \and Michel Raynal~\footnote{Institut Universitaire de France IRISA, France \texttt{michel.raynal@irisa.fr}}  \and Elad Michael Schiller~{$^\ast$}}

%
%
\maketitle              

\begin{abstract}
The problem of total-order (uniform reliable) broadcast is fundamental in fault-tolerant distributed computing since it abstracts a broad set of problems requiring processes to uniformly deliver messages in the same order in which they were sent. Existing solutions (that tolerate process failures) reduce the total-order broadcast problem to the one of multivalued consensus.

Our study aims at the design of an even more reliable solution. We do so through the lenses of \emph{self-stabilization}---a very strong notion of fault-tolerance. In addition to node and communication failures, self-stabilizing algorithms can recover after the occurrence of \emph{arbitrary transient faults}; these faults represent any violation of the assumptions according to which the system was designed to operate (as long as the algorithm code stays intact).  

This work proposes the first (to the best of our knowledge) self-stabilizing algorithm for total-order (uniform reliable) broadcast for asynchronous message-passing systems prone to process failures and transient faults. As we show, the proposed solution facilitates the elegant construction of self-stabilizing state-machine replication using bounded memory.
\end{abstract}

\Section{Introduction}
\label{sec:intro}
Fault-tolerant distributed applications span many domains in the area of banking, transport, tourism, production, and commerce, to name a few. The implementations of these applications use message-passing systems and require fault-tolerance. The task of designing and verifying these systems is known to be hard, because the joint presence of failures and asynchrony creates uncertainties about the application state (from the process's point of view). 
Our focal application is the distributed emulation of finite-state machines. For the sake of consistency maintenance, all emulating processes need to apply identical sequences of state transitions. Existing fault-tolerant solutions divide the problem into two: (i) propagation of user input to all emulating processes, and (ii) agreeing on a uniform order according to which messages are delivered. Uniform reliable broadcast~\cite{DBLP:books/sp/Raynal18,hadzilacos1994modular} can solve Problem (i). Consensus can facilitate the solution of Problem (ii).
The challenge of combining the solutions to problems (i) and (ii) is called \emph{total-order uniform reliable broadcast}~\cite{DBLP:books/sp/Raynal18}, TO-URB from now on. TO-URB lets each emulating process execute identical sequences of state transitions. 
There are fault-tolerance TO-URB implementations. 
This work aims at a more fault-tolerant TO-URB than the existing ones.

\Subsection{Problem definition} 
\label{sec:probIntro}
We study the TO-URB problem (Definition~\ref{def:URB}). It uses the operations TO-broadcast (for sending application messages) and TO-deliver (for receiving them). 




\begin{definition}
	\label{def:URB}
	TO-URB requires the satisfaction of the following. 
	\begin{itemize}
		\item \textbf{TO-validity.~~} Suppose a process TO-delivers $m$. Message $m$ was previously TO-broadcast by its sender, which is denoted by $m.sender$.
		
		\item \textbf{TO-integrity.~~} A process TO-delivers $m$ at most once.
		
		\item \textbf{TO-delivery.~~} Suppose a process TO-delivers $m$ and later TO-delivers $m'$. No process TO-delivers $m'$ before $m$.
		
		\item \textbf{TO-completion-1.~~} Suppose a non-faulty process TO-broadcasts $m$. All non-faulty processes TO-delivers $m$.
		
		\item \textbf{TO-completion-2.~~} Suppose a process TO-delivers $m$. All non-faulty processes TO-deliver $m$.

	\end{itemize}
\end{definition}



	\begin{figure}
		\begin{center}
			\includegraphics[scale=0.4, clip]{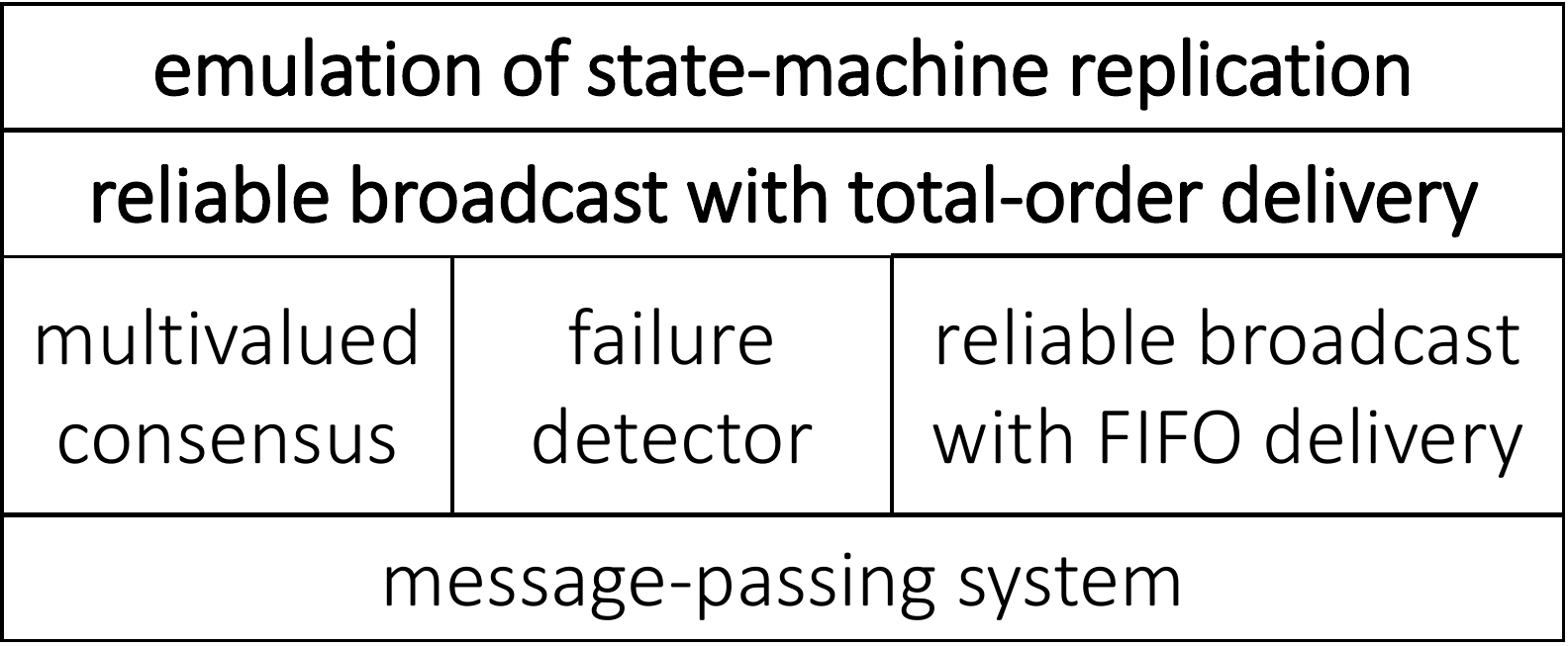}
		\end{center}
		\caption{\label{fig:suit}{\small{The context of the studied problems, ,which appear in bold font}}}
	\end{figure}

\Subsection{The studied problems and their architectural context} 
\label{sec:arc}
It is known that TO-broadcast's implementation requires the computability power of consensus, but FIFO-URB does not, see Raynal~\cite{DBLP:books/sp/Raynal18}. 
Thus, our reference architecture (\Figure~\ref{fig:suit}) includes consensus (specified in Section~\ref{sec:consensus}) and a failure detector for eventually identifying faulty nodes (Section~\ref{sec:arbitraryNodeFaults}).
It also uses the communication abstraction of FIFO-URB, which is simpler than TO-URB since it does not require the computability power of consensus. 
One can specify FIFO-URB (Section~\ref{sec:fifoURB}) by substituting the TO-delivery requirement of Definition~\ref{def:URB} with the following FIFO-delivery requirement. 
Suppose a process FIFO-delivers $m$ and later FIFO-delivers $m'$, such that the sender of these messages is the same, \ie $m.sender=m'.sender$. Then, no process FIFO-delivers $m'$ before $m$.

\Subsection{Fault model}
\label{sec:fm}
We study an asynchronous message-passing system that has no guarantees on the communication delay and the algorithm cannot explicitly access the local clock. 
We assume that this asynchronous system is prone to (detectable) fail-stop failures after which the failed node stops taking steps forever. We also consider communication failures, \eg packet loss, duplication, and reordering, as long as fair communication holds, \ie a message that is sent infinitely often is received infinitely often. We say that the faults above are \emph{foreseen} since they are known at the design time. 

In addition, we consider \emph{(arbitrary) transient faults}, i.e., any temporary violation of assumptions according to which the system was designed to operate, e.g., state corruption due to soft errors. 
We assume that these transient faults arbitrarily change the system state in unpredictable manners (while keeping the program code intact). 
We say that these faults are \emph{unforeseen} since their exact impact is unknown \tr{ at the design time}. 
In practice, a distributed system can have a non-trivial set of unknown faults that are hard to observe due to their transient nature, and thus, they cannot be individually specified as part of the fault model.

\Subsection{Design criteria}
\label{sec:designCriteria}
We aim at assuring that (if no unforeseen failures ever occur) the system, always, remains in a correct state. 
\Ie the system satisfies the task requirements, under the assumption that it starts in a correct state and that its state changes only due to algorithmic steps and foreseen failures. 
Arora and Gouda~\cite{DBLP:journals/tse/AroraG93} refer to this as the \emph{Closure} property.  

The Closure property is unattainable in the presence of unforeseen failures. 
To address such concerns, we consider a design criterion that requires the eventual system recovery (in the presence of all foreseen failures) after the occurrence of the last unforeseen and transient failure. 
Arora and Gouda~\cite{DBLP:journals/tse/AroraG93} refer to this requirement as the \emph{Convergence} property.
In other words, our design criteria require the correctness proof to demonstrate Closure and Convergence. 
In detail:
\begin{itemize}
		\item Since unforeseen failures are rare, it is assumed that all transient faults occurred before the start of the system run. 	
		\item As mentioned, transient faults can corrupt the entire system state. 
		Thus, starting from an arbitrary state, Convergence is demonstrated in the presence of foreseen failures (while assuming that the last unforeseen failure has already occurred) without the need to show that the system satisfies the task requirements. 
		\item Also, if unforeseen failures had never occurred (or after Convergence is done), the Closure property is demonstrated, i.e., the system satisfies the task requirements under the assumption that, starting from a legitimate state, the system state changes only due to the algorithmic steps and the foreseen failures.
	\end{itemize}

\Subsection{Self-stabilizing systems}
\label{sec:dc}
Dijkstra~\cite{DBLP:journals/cacm/Dijkstra74} requires self-stabilizing systems, which may start in any state, to return to correct behavior eventually. 
\Ie within a finite period, Convergence is done, and Closure is never violated. 

\Subsubsection{Asynchronous systems in the presence of stale information}
Asynchronous systems (with bounded memory and channel capacity) can indefinitely hide stale information that transient faults introduce unexpectedly. 
At any time, this corrupted data can cause the system to violate safety requirements, e.g., data consistency might be lost. 

\Ie the adversarial scheduler can both (i) violate liveness guarantees, \eg defer the task completion, and (ii) use a bounded number of opportunities to disrupt the system via a systematic exposure of hidden stale information. 
The timing of these exposures can aim at prolonging (and, if possible, preventing) recovery from the last occurrence of a transient fault. 

Due to such reasons, self-stabilizing systems often assume fair scheduling, i.e., nodes that have applicable steps (infinitely often) are allowed to take any step eventually. 
This allows self-stabilizing systems to remove, within a bounded time, all stale information whenever they appear. 
\Ie Convergence is done within a bounded time after which Closure always holds.

\Subsubsection{Asynchronous systems without any fairness assumptions}
\label{ln:wOaFa}

\noindent
Without any kind of fairness assumptions, some elementary problems do not have a straightforward answer. 
For example, a transient fault can cause a bounded counter to reach its maximum value, and yet the system might need to increment the counter an unbounded number of times after that overflow event. 
There are cases in which there is no elegant way to maintain order among the different counter values, say, by wrapping around to zero upon counter overflow. 
Thus, without any assumption on fair scheduling, a system that takes an extraordinary (or even an infinite) number of steps is bound to break any ordering constraint, because the scheduler can arbitrarily suspend node operations and defer message arrivals until such violations occur. 
Having practical systems in mind, we consider this number of (sequential) steps to be no more than practically infinite~\cite{DBLP:journals/jcss/AlonADDPT15,DBLP:journals/jcss/DolevGMS18,DBLP:journals/jcss/DolevKS10,DBLP:conf/netys/SalemS18}, say, $2^{64}$, since sequentially counting from zero to $2^{64}$ takes longer than the system's practical lifetime. 
For example, assuming a message is sent or received every nanosecond, counting from zero to $2^{64}$ takes more than $580$ years.

\Subsection{Self-stabilizing systems in the presence of seldom fairness}
\label{sec:dcA}
Dolev, Petig, and Schiller~\cite{DBLP:journals/corr/abs-1806-03498} studied self-stabilizing systems that their scheduler is seldom fair. Specifically, after the occurrence of the last transient fault, fairness eventually holds, but only for the bounded period that is sufficient for \tr{enabling} Convergence. Note that, in the absence of transient faults (or after Convergence is done), Closure is demonstrated without any fairness assumptions. Since transient faults are rare, our fairness assumption is seldom needed.  

\Subsection{Related work}
\label{sec:rw}
Non-self-stabilizing fault-tolerant TO-URB exists~\cite{DBLP:books/sp/Raynal18,hadzilacos1994modular}, but we are interested in self-stabilizing solutions. 
Seldom fairness was used for solving self-stabilizing FIFO-URB~\cite{DBLP:conf/netys/LundstromRS20}, binary and multivalued consensus~\cite{DBLP:conf/icdcn/LundstromRS21,DBLP:conf/edcc/LundstromRS21}, atomic shared memory emulation and their wait-free snapshots~\cite{DBLP:journals/corr/abs-1806-03498,DBLP:conf/netys/GeorgiouLS19}, as well as set-constraint broadcast~\cite{DBLP:conf/icdcs/LundstromRS20}, to name a few. 
This earlier literature assumes seldom fairness and shows how to transform a non-self-stabilizing algorithm into a self-stabilizing one. This work uses some of these solutions~\cite{DBLP:conf/netys/LundstromRS20,DBLP:conf/icdcn/LundstromRS21,DBLP:conf/edcc/LundstromRS21} as external building blocks. 
We note that making one set of assumptions in the absence of transient faults and then another set of assumptions in their presence is also used in the context of self-stabilizing Byzantine-fault tolerance~\cite{DBLP:journals/corr/abs-2110-08592,DBLP:journals/corr/abs-2201-12880,DBLP:conf/netys/GeorgiouMRS21}. Also, we are not the first to use self-stabilizing unreliable failure detectors~\cite{DBLP:conf/netys/BlanchardDBD14,DBLP:journals/jcss/CaniniSSSS22,DBLP:conf/netys/DolevGMS17,DBLP:journals/jcss/DolevGMS18}.


Dolev \etal~\cite{DBLP:journals/jcss/DolevGMS18} proposed a practically-stabilizing state machine replication via virtual synchrony. Note that practically-self-stabilizing systems cannot guarantee Convergence within a finite time whereas the proposed solution does. \tr{Also, the techniques in use, \ie virtual synchrony and consensus, are not identical. We note that the same holds for all related practically-stabilizing systems~\cite{DBLP:journals/jcss/AlonADDPT15,DBLP:conf/podc/BonomiDPR15,DBLP:conf/netys/SalemS18}.}      
Recently, Johnen, Arantes, and Sens~\cite{DBLP:conf/srds/JohnenA021} proposed a non-self-stabilizing yet bounded FIFO-URB and TO-URB. Our proposal is both bounded and self-stabilizing.

We note the existence of self-stabilizing systems that tolerate Byzantine behavior~\cite{DBLP:journals/tcs/BonomiPP18,DBLP:conf/srds/BonomiPPT17,DBLP:journals/tcs/BonnetDNP16,DBLP:conf/podc/BonomiPPT16,DBLP:conf/srds/MaurerT14}. Such systems are outside the scope of our fault model since they often require other kinds of solutions. For example, Dolev \etal~\cite{DBLP:conf/cscml/DolevGMS18} used partial synchrony for self-stabilizing Byzantine fault tolerant emulation of state machine replication. Also, Georgiou \etal~\cite{DBLP:conf/netys/GeorgiouMRS21} provide a self-stabilizing Byzantine fault-tolerant solution for binary consensus using randomization. The proposed solution is deterministic and does not consider partial synchrony.

It is well-known that self-stabilizing systems cannot stop sending messages when the system's task has so-called  ``terminated'', see~\cite[Chapter 2.3]{DBLP:books/mit/Dolev2000} for details. This impossibility is, mistakenly, stated as ``self-stabilizing system can never terminate''. However, the system's task can terminate but the system cannot stop sending messages. To avoid this confusion, we use the term completion rather than termination, which is the term that often appears in the literature.

\Subsection{Our contribution}
\label{sec:cont}
We present a fundamental module for dependable distributed systems: a self-stabilizing fault-tolerant TO-URB for asynchronous message-passing systems. 
Our solution assumes the availability of self-stabilizing algorithms for FIFO-URB and multivalued consensus. In the absence of transient faults, our asynchronous solution for self-stabilizing TO-URB completes within a constant number of communication rounds. After the occurrence of the last transient fault, the system recovers eventually (while assuming execution fairness among the non-faulty processes). The amount of memory used by the proposed algorithm as well as its communication costs are bounded.
To the best of our knowledge, we propose the first self-stabilizing TO-URB solution.

%

\Section{System Settings}
%
\label{sec:sys} 
We focus on asynchronous message-passing systems that have no guarantees on the communication delay. Also, the algorithm cannot explicitly access the \tr{ (local)} clock (or \tr{ use} timeout mechanisms). The system consists of a set, $\sP$, of $n$ fail-prone nodes \tr{ (or processes)} with unique identifiers. Any pair of nodes $p_i,p_j \in \sP$ has access to a bidirectional communication channel, $\mathit{channel}_{j,i}$, that, at any time, has at most $\capacity \in \bZ^+$ messages on transit from $p_j$ to $p_i$ (this assumption is due to a known impossibility~\cite[Chapter 3.2]{DBLP:books/mit/Dolev2000}). 

\emsT{When referring to an object $x$, say a variable or a function, that the state of $p_i \in \sP$ includes, and respectively, $p_i$ executes, we write $x_i$, and respectively, $x_i()$. \Ie $x$ serves as the object (variable or field) name and $x()$ is the function name. Also, when writing $x_i$, we refer to $p_i$'s storage of variable $x$ and $x_i()$ is $p_i$'s invocation of function $x()$.}
	
%
\label{sec:interModel}
In the \emph{interleaving model}~\cite{DBLP:books/mit/Dolev2000}, the node's program is a sequence of \emph{(atomic) steps}. Each step starts with an internal computation and finishes with a single communication operation, \ie a message $send$ or $receive$. The \emph{state}, $s_i$, of node $p_i \in \sP$ includes all of $p_i$'s variables and $\mathit{channel}_{j,i}$. The term \emph{system state} (or configuration) refers to the tuple $c = (s_1, s_2, \cdots,  s_n)$. We define an \emph{execution (or run)} $R={c[0],a[0],c[1],a[1],\ldots}$ as an alternating sequence of system states $c[x]$ \emsT{(configurations)} and \emsT{(atomic)} steps $a[x]$, such that each $c[x+1]$, except for the starting one, $c[0]$, is obtained from $c[x]$ by $a[x]$'s execution. \emsT{Note that we use the index, $x$, for the system states ($c[x]$) and steps ($a[x]$).}


\Subsection{The fault model and self-stabilization}
\label{sec:fMsS}
The \emph{legal executions} ($LE$) set refers to all the executions in which the requirements of task $T$ hold. \tr{In this work, $T_{\text{TO-URB}}$ denotes the task of total-order uniform reliable broadcast, which Definition~\ref{def:URB} specifies, and the executions in the set $LE_{\text{TO-URB}}$ fulfill $T_{\text{TO-URB}}$'s requirements.}

\Subsubsection{Benign failures}
\label{sec:benignFailures}
A failure occurrence is a step that the environment takes rather than the algorithm. When the failure occurrence cannot cause the system execution to lose legality, \ie to leave $LE$, we refer to that failure as a benign one.

\Subsubsubsection{Communication failures and fairness}
We focus on solutions that are oriented towards asynchronous message-passing systems and thus they are oblivious to the time at which the packets depart and arrive. We assume that any message can reside in a communication channel only for a finite period. Also, the communication channels are prone to packet failures, such as loss, duplication, and reordering.  However, if $p_i$ sends a message infinitely often to $p_j$, node $p_j$ receives that message infinitely often. This is called the \emph{fair communication} assumption. The correctness proof uses Assumption~\ref{thm:messageArrival}.

	\begin{assumption}
		\label{thm:messageArrival}
		Any sent message arrives or is lost within $\bigO(1)$ asynchronous cycles. Any URB message arrives within $\bigO(1)$ asynchronous cycles~\cite{selfStabURB}. Each active multivalued consensus object decides within $\bigO(1)$ asynchronous cycles~\cite{DBLP:conf/icdcn/LundstromRS21}.
	\end{assumption}

%
%


\Subsubsubsection{Fail-stop node failures}
\label{sec:arbitraryNodeFaults}
The system is prone to (detectable) \emph{fail-stop failures}, in which nodes stop taking steps forever. We assume at most $t<n/2$ node may fail. Denote by $\Correct$ the set of indices of nodes that never fail. We assume the availability of a self-stabilizing failure detector, such as the one by Beauquier and Kekkonen{-}Moneta~\cite{DBLP:journals/ijsysc/BeauquierK97} or Blanchard \etal~\cite{DBLP:conf/netys/BlanchardDBD14}. The interface to the failure detector offers the register $\mathit{trusted}$ that stores the local set of indexes of all nodes that are currently not suspected of being faulty.

\Subsubsection{Arbitrary transient faults}
\label{sec:arbitraryTransientFaults}
We consider any temporary violation of the assumptions according to which the system was designed to operate. We refer to these violations and deviations as \emph{arbitrary transient faults} and assume that they can corrupt the system state arbitrarily (while keeping the program code intact). The occurrence of a transient fault is rare. Thus, we assume that the last arbitrary transient fault occurs before the system execution starts~\cite{DBLP:books/mit/Dolev2000}. Also, it leaves the system to start in an arbitrary state.

\Subsection{Dijkstra's self-stabilization}
\label{sec:Dijkstra}
An algorithm is \emph{self-stabilizing} with respect to $LE$, when every execution $R$ of the algorithm reaches within a finite period a suffix $R_{legal} \in LE$ that is legal~\cite{DBLP:series/synthesis/2019Altisen,DBLP:books/mit/Dolev2000}. Namely, Dijkstra~\cite{DBLP:journals/cacm/Dijkstra74} requires $\forall R:\exists R': R=R' \circ R_{legal} \land R_{legal} \in LE \land |R'| \in \bZ^+$, where the operator $\circ$ denotes that $R=R' \circ R''$ is the concatenation of prefix $R'$ with suffix $R''$. \ea{This work assumes execution fairness only during the period, $R'$, of recovery from the occurrence of the last arbitrary transient fault.}

\tr{The part of the proof that shows the existence of $R'$ is called the \emph{convergence}, and the part that shows that $R_{legal} \in LE$ is called the \emph{closure} proof. The main complexity measure of a self-stabilizing system is the length of the recovery period, $R'$, which is counted by the number of its asynchronous communication rounds during fair executions, as we define in Section~\ref{sec:asynchronousRounds}.
} 

\tr{
	\Subsection{Execution fairness} 
\label{sec:fairnessEx}
This work assumes execution fairness only during the period in which the system recovers from the occurrence of the last arbitrary transient fault. Given a step $a$, we say that $a$ is \emph{applicable} to system state $c$ if there exists system state $c'$, such that $a$ leads to $c'$ from $c$. We say that a system execution is \emph{fair} when every step of a correct node that is applicable infinitely often is executed infinitely often, and fair communication is kept.
} 

%


\tr{
\Subsubsection{Asynchronous communication cycles}
\label{sec:asynchronousRounds}
Self-stabilizing algorithms cannot (terminate their execution and) stop sending messages~\cite[Chapter 2.3]{DBLP:books/mit/Dolev2000}. Their code includes a do-forever loop. The main complexity measure of a self-stabilizing system is the length of the recovery period, $R'$, which is counted by the number of its \emph{asynchronous cycles} during fair executions. The first asynchronous cycle $R'$ of execution $R=R'\circ R''$ is the shortest prefix of $R$ in which every correct node executes one complete iteration of the do forever loop and completes one round trip with every correct node that it sent messages to during that iteration. The second asynchronous cycle of $R$ is the first asynchronous cycle of $R''$ and so on. 
} 

\begin{figure}
	\begin{center}
		\includegraphics[scale=0.4, clip]{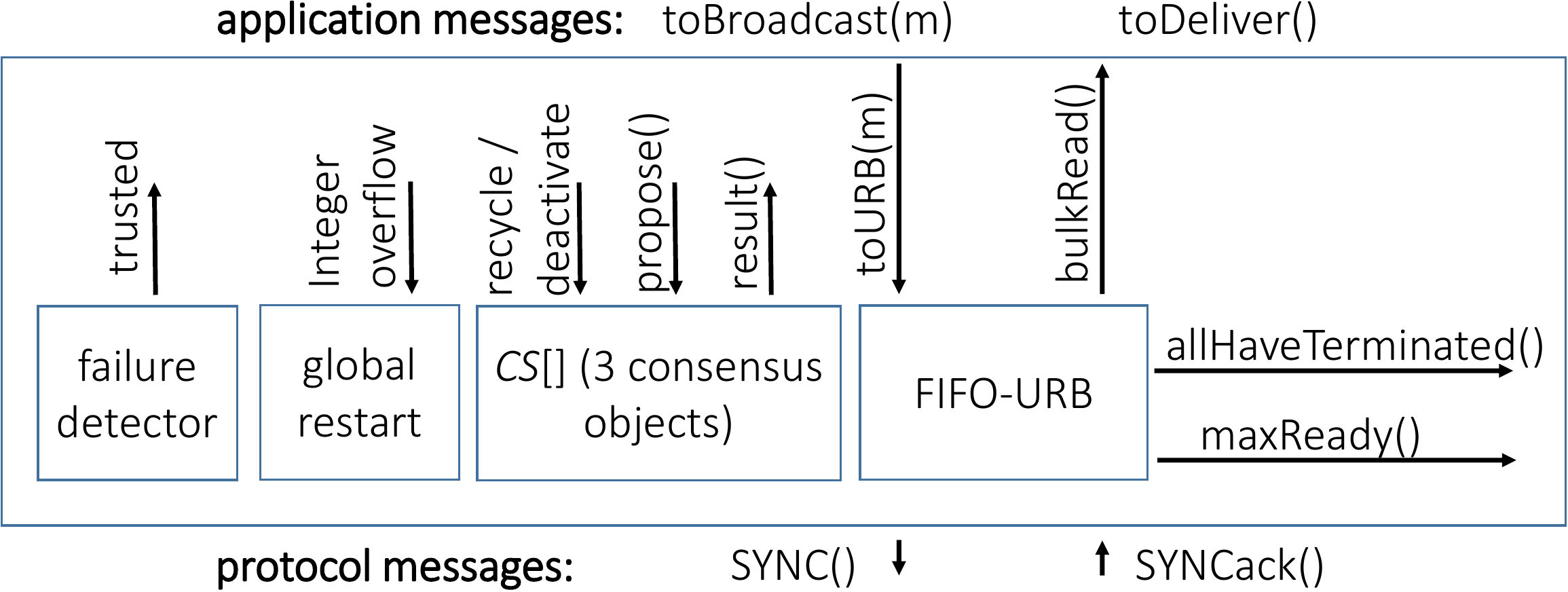}
	\end{center}
	\caption{\small{\label{fig:flow}{Info. flow among Algorithm~\ref{alg:urbTO}'s components. The proposed solution's operations, $\mathsf{toBroadcast}()$ and $\mathsf{toDeliver}()$, deal with the application messages and the solution itself coordinates its actions via protocol messages, \ie $\SYNC()$ and $\SYNCack()$.}}}
\end{figure}

\Subsection{External building blocks} 
\label{sec:ext}
The proposed solution uses a number of self-stabilizing modules (\Figure~\ref{fig:flow}). As mentioned, we assume the availability of a self-stabilizing failure detector (Section~\ref{sec:arbitraryNodeFaults}). We also assume the use of the following building blocks.

\Subsubsection{Global restart} 
\label{sec:restart}
In order to overcome the integer overflow problem (Section~\ref{ln:wOaFa}), use a global restart mechanism~\cite[Section~5]{DBLP:conf/netys/GeorgiouLS19} for initializing the system state whenever an overflow occurs. 
We assume that the maximum value in these integers is practically infinite, say, $2^{64}-1$. 
%
%
Also, in the event of integer overflow, the system runs a global restart procedure after which it cannot overflow again before it has taken $2^{64}$ communication rounds. 
We assume that no practical setup allows the system to take so many steps during its lifetime.  

\Subsubsection{Multivalued consensus} 
\label{sec:consensus}
This work uses the multivalued version of the Consensus problem (Definition~\ref{def:consensus}). \ea{Existing self-stabilizing solutions include the one by Lundstr{\"{o}}m, Raynal, and Schiller~\cite{DBLP:conf/edcc/LundstromRS21}.} \tr{ Note that there is another version of the problem in which this set includes exactly two values, and is referred to as binary consensus. Existing self-stabilizing solutions for the binary and multivalued consensus include the ones by Lundstr{\"{o}}m, Raynal, and Schiller~\cite{DBLP:conf/netys/LundstromRS20,DBLP:conf/edcc/LundstromRS21}.}

\begin{definition}[Consensus]
	\label{def:consensus}
	Every process $p_i$ has to propose a value $v_i \in V$ via an invocation of the $\mathsf{propose}_i(v_i)$ operation, where $V$ is a finite set of values. We say that algorithm $\mathit{Alg}$ solves consensus if it satisfies:
	\begin{itemize}
		\item \textbf{Validity.} Suppose that $v$ is decided. At least one process invoked $\mathsf{propose}(v)$.
		\item \textbf{Termination.} All non-faulty processes decide.
		\item \textbf{Agreement.} No two processes decide on different values.
		\item \textbf{Integrity.} No process decides more than once.
	\end{itemize}
\end{definition}

For a given consensus object $x$, we say that $x$ is active if $x\neq\bot$. In order to invoke (and activate) $x$, the algorithm calls $x.propose(v)$, where $v$ is the proposed value. As long as the consensus procedure is not completed, $x.\done()$ returns $\bot$. If an error, which is internal to $x$ occurs, $x.\done()$ returns $\blitza$. The algorithm can return $x$ to its initial state by assigning $\bot$ to $x$. Whenever $x.\done()$ returns a value that is neither $\bot$ nor $\blitza$ that value satisfies the requirements in Definition~\ref{def:consensus}.  

\Subsubsection{FIFO-URB} 
\label{sec:fifoURB}
The proposed solution assumes the availability of a well-known extension to URB called FIFO-URB. We assume the availability of a self-stabilizing FIFO-URB, such as the one by Lundstr{\"{o}}m, Raynal, and Schiller~\cite{DBLP:conf/netys/LundstromRS20}. One can specify FIFO-URB by substituting the TO-delivery requirement of Definition~\ref{def:URB} with the FIFO-delivery requirement (Section~\ref{sec:arc}).


We separate data dissemination and control. The former is carried by a FIFO-URB component and the latter by the proposed algorithm. To that end, we assume that the FIFO-URB module has interface functions that can aggregate protocol messages before their delivery. 
Specifically, we assume that the interface function $\nonActive_i()$ returns $\true$ whenever there are no active URB transmissions sent by $p_i \in \sP$. Also, given $p_i \in \sP$, the functions $\minReady_i()$ and $\maxReady_i()$ return each a vector, $r_i[0,..,n\text{-}1]$, such that for any $p_j \in \sP$, the entry $r_i[j]$ holds the lowest, and respectively, highest FIFO-delivery message number that is ready-to-be-delivered. These message numbers are the unique sequence numbers that the senders attach to the URB messages. 
The function $\bulkRead()$ allows bulk read of a set of FIFO-URB messages. Specifically, suppose $\bulkRead_i(r)$ returns immediately after system state $c$, where $\forall p_k \in \sP: r[k]\leq r_{\max}[k]$ and $r_{\max}=\maxReady_i()$. Its returned value is a deterministically ordered sequence, $sqnc_i$, that includes all the messages with message numbers $\mathit{mn}[k]$ from all senders $p_k \in \sP$, such that $r_{\min}[k] \leq \mathit{mn}[k] \leq r[k]\land r_{\min}=\minReady_i()$ in $c$. 



\Section{Self-stabilizing Bounded-memory TO-URB} 

Algorithm~\ref{alg:urbTO} presents a self-stabilizing algorithm that uses bounded memory for implementing TO-URB.
It uses FIFO-URB broadcasts for disseminating the messages that were sent via TO-broadcast. 
It defers the delivery of these FIFO broadcasts (in the buffers of FIFO-URBs) until sufficient information allows all nodes to decide on their total-order.
To that end, the URB objects report the message numbers, per sender, of messages that are ready-to-be-delivered, see Section~\ref{sec:fifoURB}. By collecting these reports from the nodes, the solution can decide, via a multivalued consensus, on the set of messages that all trusted nodes are ready to deliver. 
Specifically, Algorithm~\ref{alg:urbTO} agrees on the vector of message numbers, one number per sender, that all nodes are ready to deliver their respective messages (and all earlier messages). 
Thus, the result of the agreement defines a common set of messages that all nodes are ready to deliver. 
Since the message numbers in the set are known to all nodes, one can use a straightforward deterministic total-order for delivering these buffered messages in the same order.

\Subsection{Overview of Algorithm~\ref{alg:urbTO}}
\Figure~\ref{alg:GurbTO} presents an overview of Algorithm~\ref{alg:urbTO}. Before going through the overview, we highlight its key parts.

\begin{enumerate}

\item Upon the invocation of $\mathsf{toBroadcast}(m)$, disseminate the application message $m$ by using FIFO-URB for broadcasting $\mathsf{toURB}(m)$, \emsT{which is the name of the URB messages that need to be totally ordered before delivery.} 

\item Do forever

\begin{enumerate}

\item Query all trusted nodes about the system's consensus round numbers and the vector, $\mathit{allReady}_i$, of ready-to-be-delivered messages.


\item Recycle unused consensus objects; use round numbers info. from step (a). 

\item If the set of consensus round numbers (collected in line~\ref{ln:UanGetsAnPlusOne}) include just one number, continue to the next consensus round by proposing $\mathit{allReady}_i$. Once the consensus has been completed, $p_i$ delivers the buffered messages that their individual message numbers, per sender, are not greater than the respective entries in the agreed vector. 

\end{enumerate}
\end{enumerate}

\begin{figure}[t!]
\fbox{
\begin{minipage}{0.975\textwidth}
	\begin{\algSize}
			
			
			\renewcommand{\baselinestretch}{1.05}

			\smallskip
			
			\textbf{variables:} 
			$\CS[0..\6]$ $=[\bot,\bot,\bot]:$ consensus objects, where the proposed values are $(seq,ready)$, $seq$ is a consensus round number, and $ready$ is a vector of URB message numbers (one number per node)\label{ln:GvarRxSTOA}.
			
			$\xS=0:$ a local copy of the highest, possibly obsolete, consensus round number\label{ln:GvarRxSTO}. 
			
						\smallskip
			
			\textbf{macros:} 
			$\exceed()$: indicates the need for flushing the buffer, \ie all URBs have been completed, or the number of messages exceeds a predefined constant, $\delta$. 
			
			\begin{enumerate}
			
			\item \textbf{operation} $\mathsf{toBroadcast}(m)$ \textbf{do} \label{ln:GOptoBroadcast} \emsT{FIFO-URB the message $m$ along with the message name $\mathsf{toURB}$.} 
			
			\item \textbf{do forever} \label{ln:GdoForEver} 
			
			\begin{enumerate}
				
				\item \underline{Collect info. about round numbers and buffered messages.} Query all trusted nodes, $p_j$, about $\xS_j$, $\getSeq_j()$, which is the highest consensus round number known to $p_j$, and $\maxReady_j()$, which is a vector of $p_j$'s ready-to-be-delivered $\mathsf{toURB}()$ messages\label{ln:UanGetsAnPlusOne} (lines~\ref{ln:anGetsAnPlusOne} to~\ref{ln:letAgregate}).
				Use the arriving values for calculating:\label{ln:GletAgregate}  
				
				\begin{enumerate}
					\item $\maxSeq_i$: the greatest collected consensus round number.

					\item $\mathit{allSeq}_i$: the set of all collected consensus round numbers.

					\item $\mathit{allReady}_i$: a vector of message numbers, per sender, of the ready-to-deliver broadcasts that all nodes can perform.
					
					
				\end{enumerate}
				
				\item \underline{Recycle unused consensus objects.} Nullify $\CS[]$'s unused entries, \ie assign $\bot$ to any $\CS[k]$, for which $k \in \{0,1,\6\}$ is not one of the following\label{ln:Gk06x7minAllSeq} (line~\ref{ln:k06x7minAllSeqGet}):
				
				
				\begin{enumerate}
					\item $\xS_i \bmod \7$, but only when $\xS < \getSeq_i()$, \ie $\CS[]$'s highest consensus round number, $\getSeq_i()$, is higher than the locally highest obsolete round number, $\xS_i$. The reason is that $p_i$ still uses this entry.
					\item $\getSeq_i()$ $\bmod~ \7$ since there might be another node that is using it.
					\item $\maxSeq_i\mathit{+}1 \bmod 3$ but only when $|\mathit{allSeq}_i|=1$, \ie there is a single consensus round number. This is because one should not nullify the next entry since another node might have already started to use it.
				\end{enumerate}

				\item \underline{Agree on the delivery order.} If one collected consensus round number exists and it is time to flush the $\mathsf{toURB}()$ buffer\label{ln:GallSeqeqoneexceedmaxSeqMinReady}, \ie $\exceed()=\true$, call $\CS_i[\maxSeq_i+1~\bmod 3].propose(\maxSeq_i+1,\mathit{allReady}_i)$ (lines~\ref{ln:allSeqeqoneexceedmaxSeqMinReady} to~\ref{ln:xNeqBotLand}).
				If other nodes have higher rounds than $p_i$ or the current consensus object has completed, then\label{ln:GxNeqBotLand}:
				
				\begin{enumerate}
					\item  If the current object has been completed, deliver all messages that their individual consensus round numbers are not greater than the agreed ones\label{ln:GtoDeliverMif} (line~\ref{ln:toDeliverMif}).
					
					\item Finish the current consensus round, \ie $\xS \gets \xS+1$\label{ln:GxSGetsxSPlusOne} (line~\ref{ln:xSGetsxSPlusOne}).

				\end{enumerate}					
				
			\end{enumerate}
		\end{enumerate}
					\smallskip
					\renewcommand{\baselinestretch}{1.0}
		
		\caption{\label{alg:GurbTO}Overview of Algorithm~\ref{alg:urbTO}; code for $p_i\in\sP$}	
		
	\end{\algSize}
\end{minipage}
}
\end{figure}

%

\Subsection{Going through the overview of \Figure~\ref{alg:GurbTO}}
\label{sec:goingOverview}
The array $\CS[]$ \emsT{stores} three multivalued consensus objects\emsT{, where the proposed values are $(seq,ready)$. The field $seq$ is a round number of a multivalued consensus invocation that moderates the ordered delivery of URB messages. The field $ready$ is a vector of URB message numbers (one number per node)---each number, say $ready[j]$, moderates the URB messages sent by $p_j$.} 
The integer $\xS$ holds the consensus round number that is locally considered to be the highest one, but possibly obsolete.
Once $p_i$ delivers the messages associated with $\CS[\xS \bmod 3].\done()$ (and the earlier ones), $p_i$ considers $\xS_i$ as obsolete.
Node $p_i$ recycles $\CS[\xS \bmod 3]$ once it knows that all other trusted nodes also consider $\xS$ as an obsolete round number. 
Algorithm~\ref{alg:urbTO} uses $\CS[]$ cyclically by considering $\xS$'s value modulus three.
As explained in Section~\ref{sec:restart}, we use global reset for dealing with the event of $\xS$'s integer overflow.

Since Algorithm~\ref{alg:urbTO} defers message delivery, there is a need to guarantee that such delivery occurs eventually. To that end, the macro $\exceed()$ identifies two cases in which the buffered messages should be flushed (\ie $\exceed()$ returns $\true$): (i) the number of deferred messages exceeds a predefined constant, and (ii) there are no active URBs.

As mentioned, the invocation of $\mathsf{toBroadcast}(m)$ (line~\ref{ln:GOptoBroadcast}) leads to FIFO-URB of $\mathsf{toURB}(m)$. The do forever loop (line~\ref{ln:GdoForEver}) makes sure that these $\mathsf{toURB}()$ messages can be delivered according to an order that all nodes agree on. To that end, a query is sent (line~\ref{ln:UanGetsAnPlusOne}) to all nodes, $p_j$, about their current consensus round number, $\xS_j$, and the highest round numbers stored in $\CS[]$, $\getSeq_j()$, as well as the current status of their ready-to-deliver FIFO-URBs, \ie $\maxReady_j()$. Node $p_i$ uses the arriving and local information (line~\ref{ln:GletAgregate}) for calculating (i) the message numbers of all-nodes ready-to-deliver broadcasts, \ie $\mathit{allReady}_i$, (ii) the maximum consensus round number, $\maxSeq_i$, and (iii) the set of all consensus round numbers that $p_i$ is aware of, \ie $\mathit{allSeq}_i$. 

This information allows $p_i$ to recycle stale entries in $\CS_i[]$. Specifically, line~\ref{ln:Gk06x7minAllSeq} nullifies entries that are not used (or about to be used). Also, if there is just one collected consensus round number, \ie $|\mathit{allSeq}_i|=1$, and it is time to flush the buffer of the $\mathsf{toURB}()$ messages, as indicated by $\exceed_i()$, then $p_i$ continues to the next agreement round by proposing the pair $(\maxSeq_i+1,\mathit{allReady}_i)$. As mentioned, such agreement on the value of the vector $\mathit{allReady}$ allows all nodes to deliver, in the same order, all the messages that their message numbers, per sender $p_k$, is not greater than  $\mathit{allReady}[k]$.

If $p_i$ notices that other nodes use a higher consensus round number than its own (which implies that they have already continued to the next consensus round) or its current consensus object has been completed, $p_i$ can deliver the buffered messages (line~\ref{ln:GxNeqBotLand}). Specifically, it tests whether the current consensus object has been completed. If so, it then delivers all messages that their individual message numbers are not greater than the one agreed by the completed object (line~\ref{ln:GtoDeliverMif}). In any case, it finishes the current consensus round by incrementing the agreement round number, $\xS$ (line~\ref{ln:GxSGetsxSPlusOne}).

\Subsection{A more detailed description of Algorithm~\ref{alg:urbTO}}
\label{sec:algDesc}
%
%
Algorithm~\ref{alg:urbTO} queries all nodes about the messages that are ready-to-be-delivered (lines~\ref{ln:anGetsAnPlusOne} to~\ref{ln:letAgregate}), recycles unused consensus objects (line~\ref{ln:k06x7minAllSeqGet}), agrees on the set of messages that are ready-to-be-delivered (line~\ref{ln:allSeqeqoneexceedmaxSeqMinReady}), and delivers these messages in the same order (lines~\ref{ln:xNeqBotLand} to~\ref{ln:xSGetsxSPlusOne}). We discuss in detail each part after describing notation, constants, variables (and how to bound them), and macros. The \fbox{boxed} code lines refer to the part that deals with the removal of stale information, which we explain  in Section~\ref{sec:boundedBoxed}.

\Subsubsection{Notations, constants, variables, and macros}
%
%
\Figure~\ref{alg:urbTOA} presents the preliminaries for Algorithm~\ref{alg:urbTO}. Modulo $3$ operations are denoted by $\mathsf{opr}_\7$, \eg $x~ +_\7~ y\equiv (x+y)\bmod\7$ and $x~ -_\7~ y\equiv (x-y)\bmod\7$. The function $\text{entrywise-min}(V)$ takes the set of vectors, $V$, and returns the vector $v[]$, such that $v[k]:p_k \in \sP$ is the smallest $k$-th entry in any vector $v \in V$, \ie $v[k]=\min \{v'[k]:v'\in V\}$.

As said, $\CS[]$ holds the consensus objects \tr{ that Algorithm~\ref{alg:urbTO} accesses}. 
%
%
Algorithm~\ref{alg:urbTO} aims at aggregating URB messages and delivering them only when all transmission activities have been completed, \ie the $\nonActive()$ function returns $\true$ (Section~\ref{sec:fifoURB}). Since the number of such transmissions is unbounded, there is a need to stop aggregating after some predefined constant number of transmissions, \ie $\delta$. The variable $\xS$ points to the local (highest), possibly obsolete, consensus round number. 
The integer $\nextQuery$ stores the sequence number of the next query. As explained in Section~\ref{sec:restart}, we use global reset for dealing with the event of integer overflow for the variables $\xS$ and $\nextQuery$. 


The macro $\Sset()$ returns the set of consensus round numbers used by the active, \ie non-$\bot$ entries in $\CS[]$. The macro $\getSeq()$ returns the maximum consensus round number in $\Sset()$.
%
%
The macro $\exceed()$ facilitates the decision about whether to invoke a new consensus. It returns $\true$ if there are non-delivered messages but no ongoing transmissions, \ie $\nonActive()$ returns $\true$. It also returns $\true$ when the number of ready-to-be-delivered messages exceeds $\delta$ (regardless of the presence of active URB transmissions).

\Subsubsection{Querying (lines~\ref{ln:anGetsAnPlusOne} to~\ref{ln:letAgregate})}
Algorithm~\ref{alg:urbTO} uses a query mechanism. Each query instance is associated with a unique query number that is stored in the variable $\nextQuery$ and incremented in line~\ref{ln:anGetsAnPlusOne}. Line~\ref{ln:URBTOsendSYNC} broadcasts the synchronization query repeatedly until a reply is received from every trusted node. The query response (line~\ref{ln:URBTOackSYNC}) includes the correspondent's maximum consensus round number stored locally by any multivalued consensus object (that the macro $\getSeq()$ retrieves), the maximum possibly obsolete consensus round number (that its respective consensus object is, perhaps, no longer needed), and the latest value returned from $\maxReady_i()$. Using these responses (line~\ref{ln:URBTOsendSYNCack}), line~\ref{ln:letAgregate} aggregates the query results and store them in $\mathit{allReady}$, $\maxSeq$, and $\mathit{allSeq}$. 
Specifically, the vector $\mathit{allReady}$ includes the entry-wise minimum (per sender) for $\mathsf{toURB}()$'s message identifiers that their messages are ready-to-be-delivered at all nodes. Also, $\maxSeq$ is the maximum known consensus round number. And, the set $\mathit{allSeq}$ includes all the maximum collected consensus round numbers and obsolete consensus round numbers.            





\Subsubsection{Agreement (line~\ref{ln:allSeqeqoneexceedmaxSeqMinReady})}
The if-statement condition in line~\ref{ln:allSeqeqoneexceedmaxSeqMinReady} tests whether all trusted nodes share the same round number in $\xS$. This happens when all trusted nodes, $p_j \in \sP$, have $\xS_j=\getSeq_j()$. Line~\ref{ln:allSeqeqoneexceedmaxSeqMinReady} also checks whether $\exceed()$ indicates that it is the time to deliver. If this is the case, then line~\ref{ln:CSseqMpropose} proposes to agree on the pair $(\xS,\mathit{allReady})$ using object $\CS[\xS+1\bmod 3]$. 

\Subsubsection{Message delivery (lines~\ref{ln:xNeqBotLand} to~\ref{ln:xSGetsxSPlusOne})}
The delivery of a new $\mathsf{toURB}()$ message batch is possible once consensus has been achieved  (line~\ref{ln:xNeqBotLand}). Before the delivery (line~\ref{ln:toDeliverMif}), there is a need to check that the consensus has been completed correctly (line~\ref{ln:toDeliverMif}), cf. the interface details in Section~\ref{sec:consensus}. In any case, $\xS$ is incremented (line~\ref{ln:xSGetsxSPlusOne}) so that even if an error occurred, the object is recycled.

\begin{figure}[t!]
\fbox{\begin{minipage}{0.97\textwidth}
	\begin{\algSize}
		\smallskip
		
		\noindent \textbf{notations:}\label{ln:xMod} $x~ \mathsf{opr}_\7~ y \equiv (x~ \mathsf{opr}~ y) \bmod \7:\mathsf{opr} \in \{\text{-},\text{+}\}$, \eg $x~ +_\7~ y\equiv (x+y)\bmod\7$\; 
		$\text{entrywise-min}(V)\equiv [x_0,\ldots,x_k,\ldots,x_{n-1}] :  x_k =\min \{v[k] :v \in V\}$.
		
		
		\smallskip
		
		\noindent \textbf{constants:} 
		$\delta \in \mathbb{Z}^+$ max number of messages after which delivery is enforced\label{ln:varRxSTOH}.
		
		\smallskip
		
		\noindent \textbf{variables:} 
		$\CS[0..\6]$ $=[\bot,\bot,\bot]:$ array of multivalued consensus objects, where the proposed values are $(seq,ready)$, $seq$ is an instance number of a consensus object, and $ready$ is a vector of URB message numbers (one per node)\label{ln:varRxSTOA}.
		
		$\xS=0:$ a local copy of the highest, possibly obsolete, consensus round number\label{ln:varRxSTO}. 
		
		$\nextQuery=0:$ is a query number\label{ln:varRxSTOB}. 
		
		\smallskip
		
		\textbf{required interface:} $\mathsf{fifoURB}(m)$ FIFO-broadcast operation. 
		
		$\nonActive()$ indicates that, currently, there are no active URBs.
		
		$\minReady()$ and $\maxReady()$ return each a vector, $r_i[0,..,n\text{-}1]$, such that $r[j]$ is the lowest, and resp., highest ready-to-deliver message number. 
		
		$\bulkRead(v)$ reads all locally ready-to-deliver messages that their individual message numbers, per sender $p_k$, is at most $v[k]$. 
		
		\smallskip
		
		\textbf{macros:} $\Sset()= \{\CS[k].\mathit{seq}:\CS[k]\neq\bot \}_{k \in \{0,\ldots,\6 \}}$ \texttt{// round numbers in $\CS[]$.}
		
		\smallskip
		
		$\getSeq()$ \textbf{do} \Return{$\max(\{\xS\} \cup \Sset())$\label{ln:returnSeq}} \texttt{// highest consensus round number of consensus objects stored locally.}
		
		%
		
		\smallskip
		
		$\exceed()$ \label{ln:exceed}\textbf{do} \Return{$((\nonActive() \land 0<\ell)$ $\lor \delta \leq\ell)$} \textbf{where} $(x,y,\ell)=(\minReady(),$ $\maxReady(), \sum_{p_k \in \sP}(y[k]-x[k]))$ \texttt{// indicates whether all URBs have been completed, or the number of buffered messages exceeds $\delta$.} 
		
		\smallskip
		
		\caption{\label{alg:urbTOA}Notations, constants, variables, and macros for Algorithm~\ref{alg:urbTO}}	
		
	\end{\algSize}
\end{minipage}
}
\end{figure}

\begin{algorithm}[t!]
\begin{\algSizeSmall}

\smallskip

\textbf{For notations, constants, variables, and macros see \Figure~\ref{alg:GurbTO}}

\smallskip

\textbf{operation} $\mathsf{toBroadcast}(m)$ \textbf{do} $\mathbf{fifoURB}(\mathsf{toURB}(m))$\label{ln:OptoBroadcast}\texttt{ // FIFO-URB message $m$} 

\smallskip

\textbf{do forever} \label{ln:doForEver}\Begin{
	%
	%
	%
	%
	%
	
	\lIf{\fbox{$(\exists k \in \{0,\ldots,\6 \} : \CS[k]\neq \bot \land \CS[k].\mathit{seq} \bmod \7 \neq k)\lor (\Sset()\neq \emptyset \land (\xS$} \fbox{$ >\max \Sset() \lor$ $\max \Sset() - \min\Sset() > 1))$\label{ln:SneqEmptySetSeq}}}{\fbox{$\CS \gets [\bot, \bot ,\bot]$ \label{ln:CSgetsBots}}}
	
	\smallskip
	
	$\nextQuery \gets \nextQuery+1$\label{ln:anGetsAnPlusOne}\tcc*{start query number $\nextQuery$. Repeat until all trusted nodes have replied (line~\ref{ln:URBTOackSYNC})}
	
	\Repeat{$\SYNCack(\nextQuery, \bullet)$ \emph{received from all} $p_j: j\in \mathit{trusted}$\label{ln:URBTOsendSYNCack}}
	{\lForEach{$p_j \in \sP$}{$\mathbf{send}~\SYNC(\nextQuery)~ \mathbf{to}~ p_j$\label{ln:URBTOsendSYNC}}}
	
	\smallskip
	
	
	\texttt{// $\mathit{allReady}$, $\maxSeq$, and $\mathit{allSeq}$ are vector of ready-to-deliver messages, maximum consensus round number, resp., set of all round numbers}
	
	\textbf{let} $(\mathit{allReady},\maxSeq,\mathit{allSeq})=(\text{entrywise-min}\{x\}_{(\bullet,x) \in X}$, $\max \{x\}_{(\bull,x,\bullet) \in X}$, $\cup_{(\bull,x,y,\bull) \in X}\{x,y\})$\label{ln:letAgregate} \textbf{where} $X$ is the set of messages received in line~\ref{ln:URBTOsendSYNCack}\;
	
	%
	
	
	\fbox{\textbf{let} $(x,y,z)=(\xS,\getSeq(), \maxSeq)$;\label{ln:letXYZ}}
	
	\fbox{\lIf{$\neg (x\text{+}1 = y = z \lor x = y =z \lor x = y= z\text{-}1)$\label{ln:xyzGetSeq}}{$\xS \gets \max \{x,y,z\}$\label{ln:xyzGetSeqGet}}}
	
	\smallskip
	
	\tcc{nullify entries in $\CS[]$ that are not used (or about to be)}
	
	\lForEach{$k \in \{0,\ldots,\6\} \setminus ( \{\xS \bmod \7 : \xS < \getSeq()\} \cup \{\getSeq()$ $\bmod~ \7 \}  \cup \{\maxSeq\text{+}_{\7}1 :|\mathit{allSeq}|=1\})$\label{ln:k06x7minAllSeq}}{$\CS[k] \gets \bot$\label{ln:k06x7minAllSeqGet}}
	
	\smallskip
	
	\tcc{start the next agreement round if there is just one consensus round number and it's time to flush the FIFO-URB buffer}
	
	\lIf{$(|\mathit{allSeq}|=1\land \exceed())$\label{ln:allSeqeqoneexceedmaxSeqMinReady}}
	{$\CS[\maxSeq+_{\7}1].propose(\maxSeq+1,\mathit{allReady})$\label{ln:CSseqMpropose}}
	
	\smallskip
	
	\tcc{deliver buffered messages if other nodes use a higher consensus round number or the current consensus object has been completed}
	
	\If{$\xS+ 1 = \getSeq() \land x \neq \bot  \land x.\done() \neq \bot$ \textbf{\emph{where}} $x=\CS[(\xS+_{\7}1)]$\label{ln:xNeqBotLand}}{{
			
			\smallskip
			
			\texttt{//deliver the agreed set of messages (if possible)}
			
			\If(\texttt{ // the symbol $\blitza$ denotes an internal state error}){$x.\done() \neq \blitza$}{\textbf{foreach }{$m \in \bulkRead(x.\done())$\label{ln:toDeliverMif}} \textbf{do} {$\mathsf{toDeliver}(m)$\label{ln:toDeliverM}}} 
		}
		
		\smallskip
		$\xS \gets \xS+1$\label{ln:xSGetsxSPlusOne} \texttt{// finish the current consensus round}
	}
}

\smallskip

\tcc{reply to $\SYNC()$ messages with $\getSeq()$, $\xS$, and $\maxReady()$, which is a vector of $p_i$'s ready-to-be-delivered messages}

\textbf{upon} $\SYNC(\mathit{\nextQueryJ})$ \textbf{arrival} \textbf{from} $p_j$ \textbf{do} $\mathbf{send}~\SYNCack(\mathit{\nextQueryJ}, \getSeq(), \xS, \maxReady())~ \mathbf{to}~ p_j$\label{ln:URBTOackSYNC}\;


\caption{\label{alg:urbTO}Self-stabilizing TO-URB via consensus; code for $p_i$}	

\end{\algSizeSmall}

\end{algorithm}

\Subsection{The removal of stale information}
\label{sec:boundedBoxed}
The occurrence of transient faults can introduce stale information via the corruption of the system state. 
As mentioned, the variables $\xS$ and $\nextQuery$ may overflow (due to a transient fault). This can lead to a global restart (Section~\ref{sec:restart}). Naturally, stale information can appear in the consensus objects of $\CS[]$. Therefore, line~\ref{ln:SneqEmptySetSeq} deactivates all the consensus objects, $x=\CS[k]$, whenever inconsistencies are observed. This occurs, specifically in line~\ref{ln:SneqEmptySetSeq}, if $k$ is not equal to $x$'s consensus round number modulo $\7$ or when $\xS$ is greater than any of the consensus round numbers stored in $\CS$, as well as the case in which that set includes anything but than one or two consecutive numbers. 

Lines~\ref{ln:letXYZ} to~\ref{ln:xyzGetSeqGet} consider the locally stored and query-collected consensus round numbers. Specifically, consistent values of $\xS$, $\getSeq()$, and $\maxSeq$ have to follow one of the three scenarios. (i) The locally highest obsolete consensus round number is smaller by one than the highest locally stored or collected number, \ie $\xS+1=\getSeq()=\maxSeq$. (ii) All locally stored or collected round numbers are the same, \ie $\xS=\getSeq()=\maxSeq$. (iii) The highest collected round number is higher by one than all local ones, \ie $\xS=\getSeq()=\maxSeq-1$.

\Section{Correctness Proof of Algorithm~\ref{alg:urbTO}}
\label{sec:algDescCorr}
Definition~\ref{def:consistentTO} defines Algorithm~\ref{alg:urbTO}'s legal executions.
Invariants (i) and (ii) consider consistent states, which have no stale information.
Invariant (iii) considers predicate $Pred$, which depicts a system state in which all correct nodes use only one consensus round number in all of their variables, and thus, the nodes continue to the next consensus round and the delivery of pending messages (lines~\ref{ln:allSeqeqoneexceedmaxSeqMinReady} to~\ref{ln:xSGetsxSPlusOne}).
Invariant (iii.a) investigates the case in which no $\mathsf{toBroadcast}()$ is invoked and requires the predicate $Pred$ to hold eventually. 
Invariant (iii.b) investigates the complementary case in which $\mathsf{toBroadcast}()$ is invoked infinitely often, and requires the predicate $Pred$ to hold infinitely often.

\begin{definition}[Consistent states and legal executions]
	\label{def:consistentTO}
	Let $c$ be a system state and $p_i \in \sP$ be any node in the system. Suppose that in $c$, it holds that (i) the if-statement condition in line~\ref{ln:SneqEmptySetSeq} holds.
	%
	%
	Moreover, (ii.a) $\nextQuery_i$'s value is greater than or equal to any $\mathit{\nextQueryJ}$ field in the message $\mathsf{SYNC}(\mathit{\nextQueryJ})$ in a communication channel from $p_i$ as well as $\mathrm{SYNCack}(\mathit{\nextQueryJ},\bullet)$ message in a communication channel to $p_i$. And (ii.b) $\xS_i \leq \getSeq_i() \leq \xS_i+\5$. In this case, we say that $c$ is consistent concerning Algorithm~\ref{alg:urbTO}. 
	
	Suppose that $R$ is an execution of Algorithm~\ref{alg:urbTO}, such that every $c \in R$ is consistent. In addition, (iii.a) suppose that if $\mathsf{toBroadcast}()$ is not invoked during $R$ nor do any FIFO-broadcast becomes available for delivery, then the predicate $pred$ holds throughout $R$, where $pred \equiv \exists z \in \mathbb{Z}^+:\forall k \in \mathit{Correct}: \getSeq_k()=z \land \mathit{maxSeq}_k=z \land \xS_k=z\land \mathit{allSeq}_k=\{z\}$. Furthermore, (iii.b) suppose that if $\mathsf{toBroadcast}()$ is invoked during $R$ infinitely often, then $pred$ holds infinitely often. In this case, we say that $R$ is legal.
\end{definition}

Theorem~\ref{thm:converTO} uses Definition~\ref{def:consistentTO} for showing that Algorithm~\ref{alg:urbTO} is a self-stabilizing implementation of TO-URB. Its proof gives both the high-level proof overview and the exact proof arguments.

\begin{theorem}
	\label{thm:converTO}
	Within $\bigO(1)$ asynchronous cycles, Algorithm~\ref{alg:urbTO}'s execution is legal w.r.t. TO-URB.
	%
\end{theorem}
\renewcommand{\thmcnt}{\ref{thm:converTO}}

\begin{theoremProof}	
	Due to line~\ref{ln:CSgetsBots}, Definition~\ref{def:consistentTO}'s Invariant (i) holds after $p_i$ first complete iteration of the do-forever loop (lines~\ref{ln:doForEver} to~\ref{ln:xSGetsxSPlusOne}). (See Section~\ref{sec:restart} for dealing with the case of $\xS$'s integer overflow event.) Lemma~\ref{thm:sn} demonstrates Invariant (ii.a) by showing that, eventually,  Algorithm~\ref{alg:urbTO} lets $p_i$ introduce a $\nextQuery_i$'s value that did not exist in the system. This value overtakes any stale information associated with $\nextQuery_i$.  
	%
	%
	Line~\ref{ln:xyzGetSeqGet} implies Invariant (ii.b). 	Lemma~\ref{thm:Q} shows invariant (iii). 
	
	\begin{lemma}
		\label{thm:sn}
		Invariant (ii.a) holds.
		\remove{Let $\nextQuery_i=x$ be the value of $p_i$'s $\nextQuery$ value in $R$'s starting system state. Within $\bigO(1)$ asynchronous cycles, $\nextQuery_i$'s value is greater than or equal to $x$ and any $\mathit{\nextQueryJ}$ field in the message $\mathsf{SYNC}(\mathit{\nextQueryJ})$ in a communication channel from $p_i$ as well as $\mathrm{SYNCack}(\mathit{\nextQueryJ},\bullet)$ message in a communication channel to $p_i$.} 
	\end{lemma}
	\renewcommand{\lemcnt}{\ref{thm:sn}}
	
	\begin{lemmaProof}
		Only line~\ref{ln:anGetsAnPlusOne} modifies $\nextQuery_i$'s value, \ie by increasing $\nextQuery_i$.
		If $R$ includes the invocation of the global restart procedure (Section~\ref{sec:restart}), then by the end of that procedure (which occurs within $\bigO(1)$ asynchronous cycles), Invariant (ii.a) holds. 
		Otherwise, $R$ does not include an integer overflow event. 
		By Assumption~\ref{thm:messageArrival}, within $\bigO(1)$ asynchronous cycles, any message is either delivered or lost. Therefore, within $\bigO(1)$ asynchronous cycles, the system includes only messages with $\nextQuery_i$'s values that line~\ref{ln:anGetsAnPlusOne} introduced to the system. 
		Thus, within $\bigO(1)$ asynchronous cycles, Invariant (ii.a) holds since $\nextQuery_i$'s value is monotonically increasing. 
		%
		%
	\end{lemmaProof}
	
	\FF
	Lemma~\ref{thm:Q}'s proof considers both invariants (iii.a), \ie Claim~\ref{thm:IiiiA}, and (iii.b), \ie Claim~\ref{thm:IiiiB}. Claim~\ref{thm:IiiiA} shows that, in the absence of $\mathsf{toBroadcast}()$ invocations, the system comes to a standstill point that allows all correct nodes to use only one consensus round number, which implies that the predicate $Pred$ holds. Claim~\ref{thm:IiiiB} has the form of a proof by contradiction and it shows that if $\mathsf{toBroadcast}()$ is invoked, infinitely often, $Pred$ eventually holds since all consensus objects complete their operations within $\bigO(1)$ asynchronous cycles.
	
	We observe from Algorithm~\ref{alg:urbTO} and Definition~\ref{def:consensus} that once invariants (i) and (ii) of Definition~\ref{def:consistentTO} hold, they are not violated. Thus, Lemma~\ref{thm:Q}, which shows invariant (iii), assumes that invariants (i) and (ii) hold in every state of $R$. 	
	
	\begin{lemma}
		\label{thm:Q}
		Within $\bigO(1)$ asynchronous cycles, $R=R'\circ R''$ reaches a suffix, $R''$, in which invariants (iii.a) and (iii.b) hold.
	\end{lemma}
	\renewcommand{\lemcnt}{\ref{thm:Q}}
	
	\begin{lemmaProof}	
		Claim~\ref{thm:M} is needed for Lemma~\ref{thm:Q}'s proof. It considers the query mechanism and shows that it collects a fresh dataset, $M_{\nextQuery_i}=\{(\nextQuery_i,s_k,o_k,r_k)\}_{k \in \mathit{trusted}_i}$, that includes one record per trusted node, where $s_k=\getSeq_k()$, $o_k=\xS_k$, and $r_k=\maxReady_k()$ are values sent by $p_k \in \sP$ after $p_i$'s current value has been assigned to $\nextQuery_i$. 
		
		\begin{claimA}
			\label{thm:M}
			Every complete iteration of the do-forever loop (lines~\ref{ln:doForEver} to~\ref{ln:xSGetsxSPlusOne}) allows $p_i \in \Correct$ to collect a fresh dataset, $M_{\nextQuery_i}$.
			%
			%
			%
			%
		\end{claimA}
		\renewcommand{\clmcnt}{\ref{thm:M}}
		\begin{claimProof}
			Due to invariant (ii.a), $\nextQuery_i$'s increment (line~\ref{ln:anGetsAnPlusOne}) creates a query number that is (associated with $p_i$ and) greater than all associated query numbers in the system. With this unique query number, the repeat-until loop (lines~\ref{ln:URBTOsendSYNC} to~\ref{ln:URBTOsendSYNCack}) gets a fresh collection of $M_{\nextQuery_i}$. This loop cannot block due to the end-condition (line~\ref{ln:URBTOsendSYNCack}), which considers only the trusted nodes. The rest of the proof is implied directly by lines~\ref{ln:anGetsAnPlusOne},~\ref{ln:URBTOsendSYNCack}, and~\ref{ln:URBTOackSYNC}. 
		\end{claimProof}
		\begin{claimA}
			\label{thm:IiiiA}
			Invariant (iii.a) holds.
		\end{claimA}
		%
		%
		\renewcommand{\clmcnt}{\ref{thm:IiiiA}}
		\begin{claimProof}	
			\textbf{Argument (1)} \emph{Within $\bigO(1)$ asynchronous cycles, the if-statement condition in line~\ref{ln:xNeqBotLand} does not hold.~~}
			
			By the assumption that no FIFO-broadcast becomes ready during $R$, it holds that the if-statement condition in line~\ref{ln:allSeqeqoneexceedmaxSeqMinReady} does not hold during $R$, because $\exceed()$ does not hold. By Assumption~\ref{thm:messageArrival}, all active multivalued consensus objects have been completed with $\bigO(1)$ asynchronous cycles. Therefore, within $\bigO(1)$ asynchronous cycles, the if-statement condition in line~\ref{ln:xNeqBotLand} cannot hold. (This is true because every time that it does hold, line~\ref{ln:xSGetsxSPlusOne} increments $\xS$, but since the if-statement condition in line~\ref{ln:allSeqeqoneexceedmaxSeqMinReady} does not hold, this can only happen once.) 
			
			\textbf{Argument (2)} \emph{Within $\bigO(1)$ asynchronous cycles, $\forall i \in \mathit{Correct}:\forall k \in \mathit{trusted}_i :\mathit{maxSeq}_i=\getSeq_k()$.~~}
			
			Due to Lemma~\ref{thm:M}, $\forall i \in \mathit{Correct}:\forall k \in \mathit{trusted}_i :\mathit{maxSeq}_i\geq\getSeq_k()$. Due to the foreach condition in line~\ref{ln:xyzGetSeq} and the if-statement in line~\ref{ln:k06x7minAllSeq}, within $\bigO(1)$ asynchronous cycles, line~\ref{ln:k06x7minAllSeqGet} deactivates any consensus object, $O_{i:p_i \in \sP,x \in \{0,\ldots, \6\}}=\mathit{CS}_i[x]$ for which $O_{i,x}.\mathit{seq}<\mathit{maxSeq}_i-1$. By using Assumption~\ref{thm:messageArrival} again, any re-activated multivalued consensus object has to complete with $\bigO(1)$ asynchronous cycles. Thus, the above implies that the state of any multivalued consensus object, active or not, does not change and that $\forall i \in \mathit{Correct}:\forall k \in \mathit{trusted}_i :\mathit{maxSeq}_i=\getSeq_k()$. 
			
			\textbf{Argument (3)} \emph{Within $\bigO(1)$ asynchronous cycles, $\xS_i = \getSeq_i()$ holds.~~}
			
			By Invariant (ii.b) of Definition~\ref{def:consistentTO}, either $\xS_i+ 1 = \getSeq_i()$ or $\xS_i = \getSeq_i()$.
			Suppose $\xS_i+ 1 = \getSeq_i()$ holds. Due $\getSeq()$'s definition as well as lines~\ref{ln:CSgetsBots} and~\ref{ln:k06x7minAllSeqGet}, $x_i \neq \bot$, where $x_i=\mathit{CS}_i[(\xS_i+_{\7}1)]$ (line~\ref{ln:xNeqBotLand}). By Assumption~\ref{thm:messageArrival}, within $\bigO(1)$ asynchronous cycles, the consensus object $x_i$ completes. Thus, the if-statement condition in line~\ref{ln:xNeqBotLand} holds and line~\ref{ln:xSGetsxSPlusOne} increments $\xS_i$ once. Therefore, $\xS_i = \getSeq_i()$ within $\bigO(1)$ asynchronous cycles.
			
			\textbf{Argument (4)} \emph{Within $\bigO(1)$ asynchronous cycles, the predicate $pred$ (Definition~\ref{def:consistentTO}) holds.~~}
			
			Since $\forall i \in \mathit{Correct}:\forall k \in \mathit{trusted}_i :\mathit{maxSeq}_i=\getSeq_k()=\xS_k$, then $\mathit{allSeq}_k=\{z\}$, where $\forall i \in \mathit{Correct}:\forall k \in \mathit{trusted}_i :z=\mathit{maxSeq}_i=\getSeq_k()=\xS_k$. Thus, $pred$ holds.
		\end{claimProof}	
		
		\begin{claimA}
			\label{thm:IiiiB}
			Argument (2) Invariant (iii.b) holds.
		\end{claimA}
		\renewcommand{\clmcnt}{\ref{thm:IiiiB}}
		\begin{claimProof}	
			Note that $\exceed_i()$ holds infinitely often by the assumption that $\mathsf{toBroadcast}()$ is invoked infinitely often and URB-completion. 
			We show that the if-statement condition in line~\ref{ln:allSeqeqoneexceedmaxSeqMinReady} holds within $\bigO(1)$ asynchronous cycles once $\exceed_i()$ holds. Suppose, towards a contradiction, $|\mathit{allSeq}|=1$ does not hold for a period longer than $\bigO(1)$ asynchronous cycles. Then, the then-statement in line~\ref{ln:CSseqMpropose} is not executed for a period longer than $\bigO(1)$ asynchronous cycles. In other words, for a period longer than $\bigO(1)$ asynchronous cycles, no new round numbers are introduced to the system. By arguments similar to the ones in Claim~\ref{thm:IiiiA}, the predicate $pred$ holds within $\bigO(1)$ asynchronous cycles. Thus, the if-statement condition in line~\ref{ln:allSeqeqoneexceedmaxSeqMinReady} holds within $\bigO(1)$ asynchronous cycles.	In other words, Invariant (iii.b) holds.
		\end{claimProof}	
	\end{lemmaProof}
\end{theoremProof}

\begin{algorithm}[t!]

\begin{\algSize}

\smallskip
\pushline\dosemic\nonl 

\medskip
\hspace*{-2.5em}\texttt{// same definitions as in \Figure~\ref{alg:GurbTO} and code line~\ref{ln:OptoBroadcast}.}\\
\medskip

\popline

\setcounter{AlgoLine}{2}

\textbf{do forever} \Begin{
	
	\pushline\dosemic\nonl 
	
	\medskip	\hspace*{-1.5em}\texttt{// same code as in lines~\ref{ln:SneqEmptySetSeq} to~\ref{ln:k06x7minAllSeqGet}.}
\medskip
	
	\popline

	\setcounter{AlgoLine}{13}
	
	\If{$(|\mathit{allSeq}|=1\land \exceed())$\label{ln:vallSeqeqoneexceedmaxSeqMinReady}}{
		$\CS[\maxSeq+_{\7}1].propose(\maxSeq+1,(state=\mathsf{getState}(),msg=\mathit{maxReady())})$\label{ln:vCSseqMpropose}
	}
	
	\If{$\xS+ 1 = \getSeq() \land x \neq \bot  \land x.\done() \neq \bot$ \textbf{\emph{where}} $x=\CS[(\xS+_{\7}1)]$\label{ln:vxNeqBotLand}}{{
			\If{$x.\done() \neq \blitza$}{$\mathsf{setState}(x.\done().state)$;
			
			\textbf{foreach }{$m \in \bulkRead(x.\done().msg)$\label{ln:vtoDeliverMif}} \textbf{do} {$\mathsf{toDeliver}(m)$\label{ln:vtoDeliverM}}; 
			
			\medskip
			\texttt{// same code as in line~\ref{ln:xSGetsxSPlusOne}.}
					
			} 
		}
	}
}

%
%

	\texttt{// same code as in line~\ref{ln:URBTOackSYNC}.}
	\medskip

\caption{\label{alg:urbTOV}Self-stabilizing state-machine replication;  code for $p_i\in\sP$}	

\end{\algSize}

\end{algorithm}

\Section{Discussion}
%
%
\label{sec:disc}
We proposed, to the best of our knowledge, the first self-stabilizing algorithm for total-order uniform reliable broadcast. This is built atop self-stabilizing algorithms for FIFO-URB and multivalued consensus. 

As an application to our proposal, Algorithm~\ref{alg:urbTOV} explains how to construct a self-stabilizing emulator for state-machine replication. Note that Algorithm~\ref{alg:urbTOV}'s line numbers are the ones of Algorithm~\ref{alg:urbTO}. Line~\ref{ln:vCSseqMpropose} of Algorithm~\ref{alg:urbTOV} proposes to agree on both the automaton state, which is retrieved by $\mathsf{getState}()$, and the bulk of FIFO-URB messages, as in line~\ref{ln:CSseqMpropose} of Algorithm~\ref{alg:urbTO}. Line~\ref{ln:vtoDeliverM} of Algorithm~\ref{alg:urbTOV} uses $\mathsf{setState}()$ for updating the local state of the automaton using the agreed state. 

We encourage the reader to use our solution and techniques when designing distributed systems that must recover from transient faults.


\begin{thebibliography}{10}
	
	\bibitem{DBLP:journals/jcss/AlonADDPT15}
	Noga Alon, Hagit Attiya, Shlomi Dolev, Swan Dubois, Maria Potop{-}Butucaru, and
	S{\'{e}}bastien Tixeuil.
	\newblock Practically stabilizing {SWMR} atomic memory in message-passing
	systems.
	\newblock {\em J. Comput. Syst. Sci.}, 81(4):692--701, 2015.
	
	\bibitem{DBLP:series/synthesis/2019Altisen}
	Karine Altisen, St{\'{e}}phane Devismes, Swan Dubois, and Franck Petit.
	\newblock {\em Introduction to Distributed Self-Stabilizing Algorithms}.
	\newblock Morgan {\&} Claypool Publishers, 2019.
	
	\bibitem{DBLP:journals/tse/AroraG93}
	Anish Arora and Mohamed~G. Gouda.
	\newblock Closure and convergence: {A} foundation of fault-tolerant computing.
	\newblock {\em {IEEE} Trans. Software Eng.}, 19(11):1015--1027, 1993.
	
	\bibitem{DBLP:journals/ijsysc/BeauquierK97}
	Joffroy Beauquier and Synn{\"{o}}ve Kekkonen{-}Moneta.
	\newblock Fault-tolerance and self-stabilization: impossibility results and
	solutions using self-stabilizing failure detectors.
	\newblock {\em Int. J. Systems Science}, 28(11):1177--1187, 1997.
	
	\bibitem{DBLP:conf/netys/BlanchardDBD14}
	Peva Blanchard, Shlomi Dolev, Joffroy Beauquier, and Sylvie Dela{\"{e}}t.
	\newblock Practically self-stabilizing {Paxos} replicated state-machine.
	\newblock In {\em {NETYS}}, volume 8593 of {\em LNCS}, pages 99--121. Springer,
	2014.
	
	\bibitem{DBLP:journals/tcs/BonnetDNP16}
	Fran{\c{c}}ois Bonnet, Xavier D{\'{e}}fago, Thanh~Dang Nguyen, and Maria
	Potop{-}Butucaru.
	\newblock Tight bound on mobile {Byzantine} agreement.
	\newblock {\em Theor. Comput. Sci.}, 609:361--373, 2016.
	
	\bibitem{DBLP:conf/podc/BonomiDPR15}
	Silvia Bonomi, Shlomi Dolev, Maria Potop{-}Butucaru, and Michel Raynal.
	\newblock Stabilizing server-based storage in {Byzantine} asynchronous
	message-passing systems: Extended abstract.
	\newblock In {\em {PODC}}, pages 471--479. {ACM}, 2015.
	
	\bibitem{DBLP:journals/tcs/BonomiPP18}
	Silvia Bonomi, Antonella~Del Pozzo, and Maria Potop{-}Butucaru.
	\newblock Optimal self-stabilizing synchronous mobile {Byzantine}-tolerant
	atomic register.
	\newblock {\em Theor. Comput. Sci.}, 709:64--79, 2018.
	
	\bibitem{DBLP:conf/podc/BonomiPPT16}
	Silvia Bonomi, Antonella~Del Pozzo, Maria Potop{-}Butucaru, and S{\'{e}}bastien
	Tixeuil.
	\newblock Optimal mobile {Byzantine} fault tolerant distributed storage:
	Extended abstract.
	\newblock In {\em {PODC}}, pages 269--278. {ACM}, 2016.
	
	\bibitem{DBLP:conf/srds/BonomiPPT17}
	Silvia Bonomi, Antonella~Del Pozzo, Maria Potop{-}Butucaru, and S{\'{e}}bastien
	Tixeuil.
	\newblock Optimal storage under unsynchronized mobile {Byzantine} faults.
	\newblock In {\em {SRDS}}, pages 154--163. {IEEE} Computer Society, 2017.
	
	\bibitem{DBLP:journals/jcss/CaniniSSSS22}
	Marco Canini, Iosif Salem, Liron Schiff, Elad~Michael Schiller, and Stefan
	Schmid.
	\newblock \emph{Renaissance}: {A} self-stabilizing distributed {SDN} control
	plane using in-band communications.
	\newblock {\em J. Comput. Syst. Sci.}, 127:91--121, 2022.
	
	\bibitem{DBLP:journals/cacm/Dijkstra74}
	Edsger~W. Dijkstra.
	\newblock Self-stabilizing systems in spite of distributed control.
	\newblock {\em Commun. {ACM}}, 17(11):643--644, 1974.
	
	\bibitem{DBLP:books/mit/Dolev2000}
	Shlomi Dolev.
	\newblock {\em Self-Stabilization}.
	\newblock {MIT} Press, 2000.
	
	\bibitem{DBLP:conf/netys/DolevGMS17}
	Shlomi Dolev, Chryssis Georgiou, Ioannis Marcoullis, and Elad~Michael Schiller.
	\newblock Self-stabilizing reconfiguration.
	\newblock In {\em {NETYS}}, volume 10299 of {\em LNCS}, pages 51--68, 2017.
	
	\bibitem{DBLP:journals/jcss/DolevGMS18}
	Shlomi Dolev, Chryssis Georgiou, Ioannis Marcoullis, and Elad~Michael Schiller.
	\newblock Practically-self-stabilizing virtual synchrony.
	\newblock {\em J. Comput. Syst. Sci.}, 96:50--73, 2018.
	
	\bibitem{DBLP:conf/cscml/DolevGMS18}
	Shlomi Dolev, Chryssis Georgiou, Ioannis Marcoullis, and Elad~Michael Schiller.
	\newblock Self-stabilizing {Byzantine} tolerant replicated state machine based
	on failure detectors.
	\newblock In {\em {CSCML}}, volume 10879 of {\em Lecture Notes in Computer
		Science}, pages 84--100. Springer, 2018.
	
	\bibitem{DBLP:journals/jcss/DolevKS10}
	Shlomi Dolev, Ronen~I. Kat, and Elad~Michael Schiller.
	\newblock When consensus meets self-stabilization.
	\newblock {\em J. Comput. Syst. Sci.}, 76(8):884--900, 2010.
	
	\bibitem{DBLP:journals/corr/abs-1806-03498}
	Shlomi Dolev, Thomas Petig, and Elad~Michael Schiller.
	\newblock Self-stabilizing and private distributed shared atomic memory in
	seldomly fair message passing networks.
	\newblock {\em Algorithmica. Also appears in CoRR}, abs/1806.03498, 2022.
	
	\bibitem{DBLP:journals/corr/abs-2110-08592}
	Romaric Duvignau, Michel Raynal, and Elad~Michael Schiller.
	\newblock Self-stabilizing byzantine- and intrusion-tolerant consensus.
	\newblock {\em CoRR}, abs/2110.08592, 2021.
	
	\bibitem{DBLP:journals/corr/abs-2201-12880}
	Romaric Duvignau, Michel Raynal, and Elad~Michael Schiller.
	\newblock Self-stabilizing byzantine-tolerant broadcast.
	\newblock {\em To appear in SSS'22 and also in CoRR}, abs/2201.12880, 2022.
	
	\bibitem{DBLP:conf/netys/GeorgiouLS19}
	Chryssis Georgiou, Oskar Lundstr{\"{o}}m, and Elad~Michael Schiller.
	\newblock Self-stabilizing snapshot objects for asynchronous failure-prone
	networked systems.
	\newblock In {\em Networked Systems, {NETYS}}, pages 113--130, 2019.
	
	\bibitem{DBLP:conf/netys/GeorgiouMRS21}
	Chryssis Georgiou, Ioannis Marcoullis, Michel Raynal, and Elad~Michael
	Schiller.
	\newblock Loosely-self-stabilizing {Byzantine}-tolerant binary consensus for
	signature-free message-passing systems.
	\newblock In {\em {NETYS}}, volume 12754 of {\em Lecture Notes in Computer
		Science}, pages 36--53. Springer, 2021.
	
	\bibitem{hadzilacos1994modular}
	Vassos Hadzilacos and Sam Toueg.
	\newblock A modular approach to fault-tolerant broadcasts and related problems.
	\newblock Technical report, Cornell Univ., Ithaca, NY, 1994.
	
	\bibitem{DBLP:conf/srds/JohnenA021}
	Colette Johnen, Luciana Arantes, and Pierre Sens.
	\newblock {FIFO} and atomic broadcast algorithms with bounded message size for
	dynamic systems.
	\newblock In {\em {SRDS}}, pages 277--287. {IEEE}, 2021.
	
	\bibitem{DBLP:conf/netys/LundstromRS20}
	Oskar Lundstr{\"{o}}m, Michel Raynal, and Elad Schiller.
	\newblock Self-stabilizing uniform reliable broadcast.
	\newblock In {\em Networked Systems {NETYS}}, volume 12129 of {\em LNCS}, pages
	296--313. Springer, 2020.
	
	\bibitem{DBLP:conf/icdcs/LundstromRS20}
	Oskar Lundstr{\"{o}}m, Michel Raynal, and Elad~Michael Schiller.
	\newblock Self-stabilizing set-constrained delivery broadcast.
	\newblock In {\em 40th {IEEE} International Conference on Distributed Computing
		Systems, {ICDCS}}, pages 617--627. {IEEE}, 2020.
	
	\bibitem{DBLP:conf/icdcn/LundstromRS21}
	Oskar Lundstr{\"{o}}m, Michel Raynal, and Elad~Michael Schiller.
	\newblock Self-stabilizing indulgent zero-degrading binary consensus.
	\newblock In {\em 22nd Distributed Computing and Networking {ICDCN}}, pages
	106--115, 2021.
	
	\bibitem{DBLP:conf/edcc/LundstromRS21}
	Oskar Lundstr{\"{o}}m, Michel Raynal, and Elad~Michael Schiller.
	\newblock Self-stabilizing multivalued consensus in asynchronous crash-prone
	systems.
	\newblock In {\em 17th European Dependable Computing Conference, {EDCC}}, pages
	111--118. {IEEE}, 2021.
	
	\bibitem{selfStabURB}
	Oskar Lundström, Michel Raynal, and Elad~M. Schiller.
	\newblock Self-stabilizing uniform reliable broadcast.
	\newblock In {\em Networked Systems, ({NETYS'20}) {Springer} {LNCS} 12129},
	pages 296--313, 2020.
	\newblock Also in {CoRR} abs/2001.03244.
	
	\bibitem{DBLP:conf/srds/MaurerT14}
	Alexandre Maurer and S{\'{e}}bastien Tixeuil.
	\newblock Self-stabilizing {Byzantine} broadcast.
	\newblock In {\em 33rd {IEEE} International Symposium on Reliable Distributed
		Systems, {SRDS}}, pages 152--160, 2014.
	
	\bibitem{DBLP:books/sp/Raynal18}
	Michel Raynal.
	\newblock {\em Fault-Tolerant Message-Passing Distributed Systems - An
		Algorithmic Approach}.
	\newblock Springer, 2018.
	
	\bibitem{DBLP:conf/netys/SalemS18}
	Iosif Salem and Elad~Michael Schiller.
	\newblock Practically-self-stabilizing vector clocks in the absence of
	execution fairness.
	\newblock In {\em Networked Systems {NETYS}}, pages 318--333, 2018.
	
\end{thebibliography}
\end{document}